\pdfoutput=1
\documentclass[11pt]{article}
\usepackage{graphicx}
\usepackage{latexsym}
\usepackage{mathrsfs}
\usepackage[cal=boondox]{mathalfa}
\usepackage[overload]{textcase}

.5 scaled \magstep4
\font\medio=cmr9.5 scaled \magstep2
\outer\def\beginsection#1\par{\medbreak\bigskip
      \message{#1}\leftline{\bf#1}\nobreak\medskip
\vskip-\parskip
      \noindent}
\begin{document}
\bibliographystyle{unsrt}

\titlepage
\vspace{1cm}
\begin{center}
{\Large {\bf The refractive index of the relic gravitons and the nHz band}}\\
\vspace{1.5 cm}
Massimo Giovannini \footnote{e-mail address: massimo.giovannini@cern.ch}\\
\vspace{1cm}
{{\sl Department of Physics, CERN, 1211 Geneva 23, Switzerland }}\\
\vspace{0.5cm}
{{\sl INFN, Section of Milan-Bicocca, 20126 Milan, Italy}}
\vspace*{1cm}
\end{center}
\vskip 0.3cm
\centerline{\medio  Abstract}
\vskip 0.5cm
If the refractive index of the relic gravitons increases during a conventional stage of inflationary evolution the spectral energy density is blue at intermediate frequencies above the fHz and then flattens out after a knee that is typically  smaller than the mHz. We investigate here the conditions leading to a sufficiently large spectral energy density in the nHz range where some peculiar signatures observed with the pulsar timing arrays have been recently attributed to cosmic gravitons. If these potential evidences are combined with the most recent bounds provided by wide-band interferometers in the audio range (i.e. between few Hz and the kHz) the allowed regions of the parameter space are compatible with both determinations and also with all the other constraints associated with the background of relic gravitons produced during inflation. The present analysis suggests that the pulsar timing arrays are sensitive to the evolution of the refractive index during 
early stages of the inflationary evolution. This physical interpretation of the preliminary empirical evidence is distinguishable from other perspectives since the high-frequency normalization, the blue spectral index and the tensor to scalar ratio cannot be independently assigned but are all related to the frequency of the knee that is ultimately determined by the competition between the rate of evolution of the refractive index and the slow-roll corrections.
\noindent
\vspace{5mm}
\vfill
\newpage
The pumping action of the space-time curvature is the primary source of relic gravitons \cite{AA1,AA1a} but when a conventional stage of inflationary expansion is replaced by a radiation-dominated epoch, the spectral energy density in critical units at the present conformal time $\tau_{0}$ (denoted hereunder by $\Omega_{gw}(\nu,\tau_{0})$) is quasi-flat for comoving frequencies $\nu$ larger than $100$ aHz \cite{AA2}. The transition across the epoch of matter-radiation equality leads to an infrared branch where $\Omega_{gw}(\nu,\tau_{0}) \propto \nu^{-2}$ between the aHz and $100$ aHz \cite{AA3}. The 
conventional inflationary evolution sems to  
exclude the possibility of a drastic deviation from scale-invariance in the direction of blue spectral indices for typical frequencies larger than fHz where, in a conservative perspective, $h_{0}^2 \Omega_{gw}(\nu,\tau_{0}) = {\mathcal O}(10^{-16.5})$. On the contrary the 
corrections typically go into the direction of a decreasing spectral slope as it happens, for 
instance, in the single-field case where, thanks to the consistency relations, the tensor spectral index $n_{T}$ is notoriously related to the tensor to scalar ratio $r_{T}$ as $n_{T} \simeq - r_{T}/8$. Since $r_{T}$ is currently assessed from the analysis of the temperature and polarization anisotropies of the Cosmic Microwave Background (CMB) it is difficult to argue that $n_{T}$ could be positive. For typical frequencies 
larger than the mHz the spectral energy density increases provided the post-inflationary evolution is stiffer than radiation. For frequencies between the fHz and the mHz the spectral index can be blue in generalized 
gravity theories and when the dominant energy condition is violated (see, for instance, \cite{CC0} for a recent review focussed on these issues). 

Gravitational waves also acquire an effective index of refraction when they travel in curved space-times \cite{CC1}. In this context the blue spectral slopes arise from a more mundane effect associated with the variation of the refractive index even if the background geometry evolves according to a conventional stage of expansion possibly supplemented 
by a standard decelerated epoch \cite{CC2}. The general idea behind this suggestion is that the effective action of single-field inflationary models involves all the different terms that include four derivatives and are suppressed by the negative powers of a large mass scale \cite{ONEW}. If parity breaking terms are included in the effective action \cite{TWO,THREE}, the relic graviton background may be polarized but this possibility has been already discussed elsewhere \cite{FOUR}. There are however non-generic models of inflation where the higher-order corrections assume a particular form since the inflaton has some particular symmetry (like a shift symmetry $\varphi \to \varphi + c$) or because the rate of inflaton roll remains
constant (and possibly larger than $1$), as it happens in certain fast-roll scenarios \cite{NON1} (see also, for instance, \cite{NON2,NON3}). Other examples include the situation where the higher-order curvature corrections are given in terms of the Gauss-Bonnet combination weighted by some inflaton dependent-coupling \cite{NON4,NON5,NON6}. In \cite{CC2} (see also \cite{CC5,CC3}) it has been argued that in all these situations the effective action of the relic gravitons is be modified and ultimately assumes the following general form:
\begin{equation}
S_{g} = \frac{1}{8 \ell_{P}^2} \int d^{4} x \biggl[ A(\tau)\, \,\partial_{\tau} h_{i j} \, \partial_{\tau} h^{i j} -  B(\tau)\, \partial_{k} h_{i j} \partial^{k} h^{i j} \biggr],
\label{ONE}
\end{equation}
where we assume that the background is conformally flat and $\partial_{i} h_{j}^{\,\,\,\, i}= h_{i}^{\,\,\,\,i} =0$. While both terms $A(\tau)$ and $B(\tau)$ depend on the conformal time coordinate $\tau$ 
we can always factor $A(\tau)$ and parametrize and introduce an effective
refractive index $n(\tau)$ associated with the interactions with the background geometry, as originally argued in Ref. \cite{CC1,CC2}):
 \begin{equation}
S_{g} = \frac{1}{8 \ell_{P}^2} \int d^{4} x \,\,a^2(\tau) \,\,\biggl[ \partial_{\tau} h_{i j} \, \partial_{\tau} h^{i j} - \frac{\partial_{k} h_{i j} \partial^{k} h^{i j}}{n^2(\tau)}  \biggr],
\label{TWO}
\end{equation}
where $a(\tau)$ is the scale factor in the conformal time coordinate; we are assuming here a conformally flat background metric $\overline{g}_{\mu\nu} = a^2(\tau) \, \eta_{\mu\nu}$. After Eq. (\ref{TWO}) has been proposed in Ref. \cite{CC2} apparently different parametrizations have been later introduced (see \cite{CC3,CC5} and discussion theierin).  The difference between these strategies consists in modifying the first term (rather than the second) inside the squared bracket of Eq. (\ref{TWO}). This choice is immaterial since the two parametrizations of the effect are related by a rescaling of the four-dimensional metric through a conformal factor that involves the refractive index itself. Equation (\ref{TWO}) simplifies by introducing a new time coordinate conventionally referred to as the $\eta$-time:
 \begin{equation}
S_{g} = \frac{1}{8 \ell_{P}^2} \int \,\,d^{4} x \,\,b^2(\eta)\,\, \biggl[ \partial_{\eta} h_{i j} \,\, \partial_{\eta} h^{i j} - \partial_{k} h_{i j} \,\,\partial^{k} h^{i j} \biggr], \qquad \qquad b(\eta) = \frac{a(\eta)}{\sqrt{n(\eta)}}, \qquad \ell_{P} = \sqrt{8 \pi G}.
\label{THREE}
\end{equation}
Since the $\eta$-time is defined by $n(\eta) \, d\eta \,=\,d\tau$ and the scale factor is everywhere continuous with its first derivative\footnote{Note that the continuity of the extrinsic curvature implies the continuity of the first (conformal) time derivative of the scale factor.}, the evolution of the refractive index is specified unambiguously by assigning $n(a)$. Equation (\ref{THREE}) generalizes the standard Ford-Parker action \cite{FIVEa} to the case of a dynamical refractive index.  Even if the phase velocity of the relic gravitons is not required to be sub-luminal we shall consider the situation where $n(a) \geq 1$. In what follows for the sake of illustration we shall be assuming that $n(a)$ changes appreciably during inflation and it goes to $1$ during the standard decelerated stage of expansion:
\begin{equation}
n(a) = n_{\ast} \frac{ (a/a_{\ast})^{\alpha} \,\,e^{- \xi (a/a_{1})}}{(a/a_{*})^{\alpha} + 1} + 1, \qquad\qquad
n_{\ast} = n_{i} (a_{\ast}/a_{i})^{\alpha} = n_{i} e^{\alpha \, N_{\ast}},
\label{NEX}
\end{equation}
where $a_{i}$ and $a_{1}$ mark, respectively, the beginning and the end of the 
inflationary epoch; $a_{*}$ defines the boundary of the refractive stage. 
What matters is not the specific analytical form (that only affects the transition regions) but the three successive physical regimes described by  Eq. (\ref{NEX}). In particular we have that $n(a) \to 1$  when $a\gg a_{1}$ 
 and the sharpness of this transition is controlled by $\xi \geq 1$. When $a_{*} < a < a_{1}$ $n(a)$ is constant (but still larger than $1$), i.e. $n(a)\simeq n_{\ast} > 1$. Finally for $a< a_{\ast}$ we have the truly refractive stage where $n (a) \simeq n_{\ast} (a/a_{\ast})^{\alpha}$. Even if in  Ref. \cite{CC2} more general possibilities have been examined (including a post-inflationary evolution of $n(a)$), for the present ends it is enough to consider the minimal case illustrated by Eq. (\ref{NEX}). 

\begin{figure}[!ht]
\centering
\includegraphics[height=6.5cm]{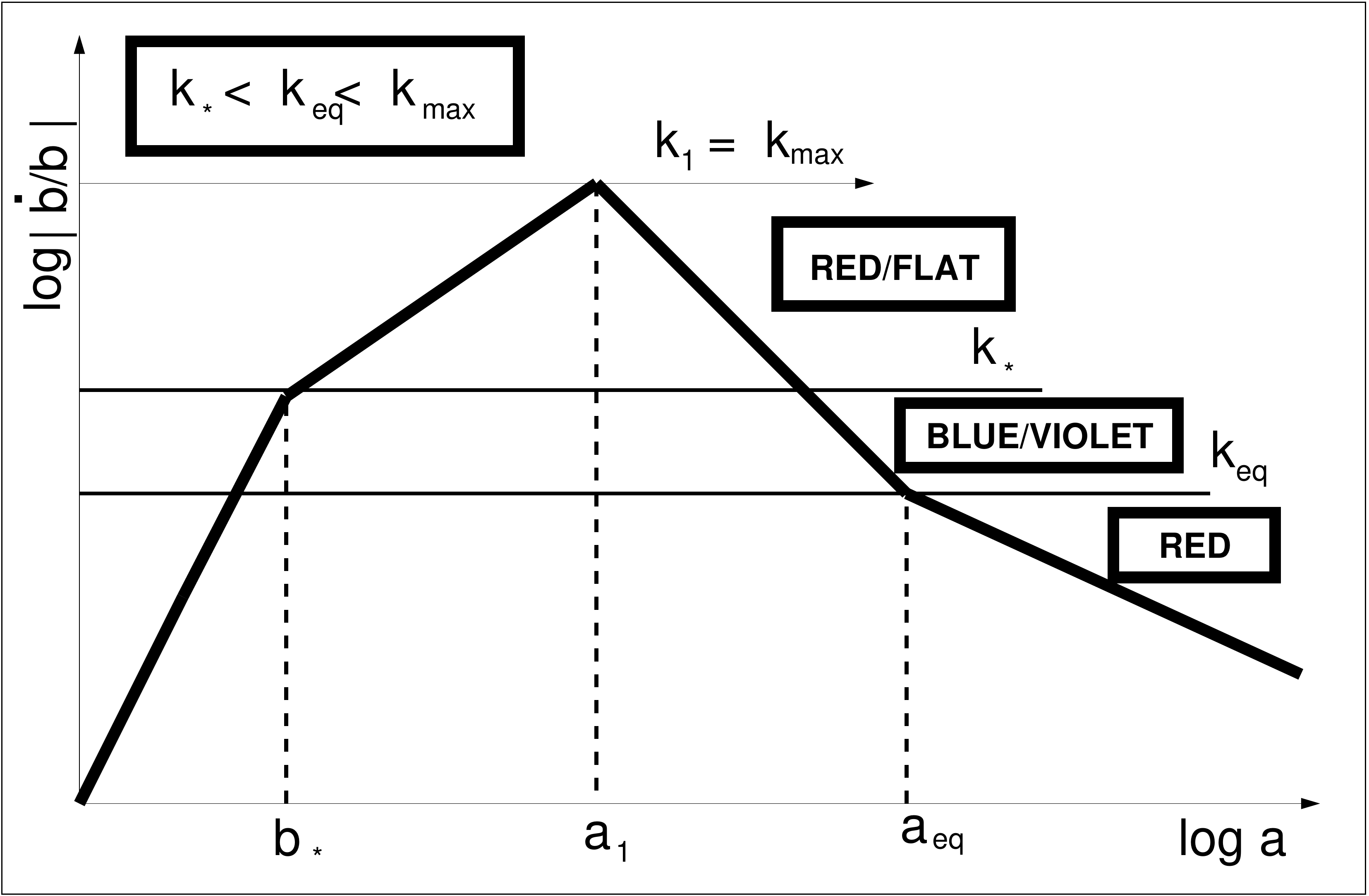}
\caption[a]{The common logarithm effective horizon is approximately illustrated in terms 
of the scale factor. The different wavenumbers hit twice the effective horizon and the two crossings 
determine the slope of the spectrum in a given dynamical stage.}
\label{FIG1}      
\end{figure}

The different dynamical stages of the model are illustrated in the cartoon of Fig. \ref{FIG1} 
where the common logarithm of the effective horizon $\dot{b}/b$ is reported. In contrast with 
the standard notations, the overdot denotes here a derivation with respect to the $\eta$ coordinate (and not a derivation with respect to the cosmic time as usually implied).  
From the shape of Fig. \ref{FIG1} we can deduce that the spectral energy density will have three different branches. The modes $ k < k_{eq}$ correspond to wavelengths leaving the Hubble radius for $b< b_{*}$  and reentering during the matter-dominated stage. We remind that, as usual, $k_{eq} = 0.0732\, h_{0}^2 \Omega_{M0}\, \mathrm{Mpc}^{-1}$ where $\Omega_{M0}$ is the present fraction in dusty matter\footnote{The comoving 
frequency associated to $k_{eq}$ is given by $\nu_{eq} = k_{eq}/(2\pi) = 1.597 \times 
10^{-17} \, \mathrm{Hz}$ for $h_{0}^2 \Omega_{M0} =0.1411$ and $h_{0}^2 \Omega_{R0}= 
4.15 \times 10^{-5}$.}.
The modes $k_{eq} < k < k_{\ast}$ (where $k_{\ast} = 1/\eta_{\ast}$) correspond to the wavelengths leaving the Hubble radius during the refractive stage and reentering when the 
background is dominated by radiation. In this range of wavenumbers the spectral energy density is either blue or violet, depending on the value of $\alpha$ which is always positive 
as long as the refractive index is larger than $1$. Finally 
in the region $ k_{\ast} < k < k_{1}$ the spectrum is quasi-flat since 
the corresponding wavelengths cross the Hubble radius during the inflationary 
phase (when the refractive index is not dynamical anymore) and reenter in 
the radiation stage of expansion. The comoving frequencies $\nu_{max}= k_{max}/(2\,\pi)$ and $\nu_{\ast}= k_{*}/(2\,\pi)$ cannot be assigned at wish depending on the convenience 
of the potential signal but they are determined from the various stages 
of the model. In particular the maximal frequency is typically ${\mathcal O}(200)$ MHz:
\begin{equation}
\nu_{max} = 0.26 \, \biggl(\frac{\epsilon}{0.003}\biggr)^{1/4} \biggl(\frac{{\mathcal A}_{\mathcal R}}{2.41 \times 10^{-9}} \biggr)^{1/4} \,  \biggl(\frac{h_{0}^2 \Omega_{R0}}{4.15 \times 10^{-5}} \biggr)^{1/4}\, \mathrm{GHz}, 
\label{NEX2}
\end{equation}
where $\epsilon$  denotes the slow-roll parameter and ${\mathcal A}_{{\mathcal R}}$ is 
the amplitude of the scalar power spectrum; according to the current 
determinations we have that the tensor to scalar ratio should be smaller than $0.064$ \cite{RT1,RT2}; in a conservative perspective we shall be requiring $\epsilon \leq 0.06/16 \simeq 0.003$ as assumed in Eq. (\ref{NEX2}).  We remind that the typical 
value of the pivot scale used in this discussion is $k_{p} = 0.002\, \mathrm{Mpc}^{-1}$
corresponding to a typical comoving frequency  $\nu_{p} =  3.092\,\,\mathrm{aHz}$.
Given $\nu_{max}$ the value of $\nu_{\ast}$ defines the knee of the spectrum:
\begin{equation}
\nu_{\ast} = \biggl(1 + \frac{\alpha}{1 - \epsilon}\biggr) e^{N_{\ast} (\alpha+1) - N_{t}} \, \, \nu_{max},
\label{NEX3}
\end{equation}
where $N_{t}$ denotes the total number of $e$-folds while, as in Eq. (\ref{NEX}),  $N_{\ast} = \ln{(a_{\ast}/a_{i})}$.  In Eq. (\ref{NEX3}) as well as in the forthcoming discussion we shall always be assuming that $n_{i} \to 1$; different choices are possible (provided $n_{i} \geq 1$) but their effect does not modify the conclusions since the value of $n_{i}$ can always be traded for a shorter refractive phase. The relevant point to appreciate is that $\nu_{\ast}$ will control the typical frequency of the knee of the spectrum and it does depend on $N_{*}$, $N_{t}$ and $\alpha$. 

The Hamiltonian associated with Eq. (\ref{THREE}) is directly expressible in the $\eta$-time and it is:
\begin{equation}
H_{g}(\eta) = \int d^{3} x \biggl[ \frac{8 \ell_{P}^2}{b^2} \pi_{i\,j} \, \pi^{\,i \, j} + \frac{b^2}{8 \ell_{P}^2} \partial_{k} h_{i\,j} \, 
\partial^{k} h^{\, i \, j} \biggr], \qquad\qquad \pi_{i\, j} = \frac{b^2}{8 \ell_{P}^2} \partial_{\eta} h_{i\,j}.
\label{THREEa}
\end{equation}
After promoting the field operators and their conjugate momenta  to the status of quantum operators, from Eq. (\ref{THREEa}) the governing equations for $\widehat{h}_{i\,j}$ and 
 $\widehat{\pi}_{i\,j}$ are:
\begin{equation}
\partial_{\eta} \widehat{h}_{i\, j} = \frac{ 8 \ell_{P}^2}{b^2}\, \widehat{\pi}_{i\, j}, \qquad \qquad \partial_{\eta} \widehat{\pi}_{i\, j} = \frac{b^2}{ 8 \ell_{P}^2}\, \nabla^2 \widehat{h}_{i\, j}.
\label{THREEb}
\end{equation}
In Fourier space Eq. (\ref{THREEb}) implies a specific solution of the corresponding 
mode functions $F_{k}(\eta)$ and $G_{k}(\eta)$; more specifically we have 
\begin{eqnarray}
\widehat{h}_{ij}(\vec{x}, \eta) &=& \frac{\sqrt{2}\, \ell_{P}}{(2\pi)^{3/2}}\sum_{\lambda} \int \, d^{3} k \,\,e^{(\lambda)}_{ij}(\vec{k})\, \biggl[ F_{k,\lambda}(\eta)\, \widehat{a}_{\vec{k},\,\lambda } e^{- i \vec{k} \cdot \vec{x}} + F^{\ast}_{k,\lambda}(\eta) \,\widehat{a}_{\vec{k},\,\lambda }^{\dagger} e^{ i \vec{k} \cdot \vec{x}}\biggr],
\label{FOURa}\\
\widehat{\pi}_{ij}(\vec{x}, \eta) &=& \frac{1}{4 \, \sqrt{2} \,\ell_{P} \, (2\pi)^{3/2}}\sum_{\lambda} \int \, d^{3} k \,\,e^{(\lambda)}_{ij}(\vec{k})\, \biggl[ G_{k,\lambda}(\eta) \,\widehat{a}_{\vec{k},\,\lambda } \,e^{- i \vec{k} \cdot \vec{x}} + G^{\ast}_{k,\lambda}(\eta) \,\widehat{a}_{\vec{k},\,\lambda }^{\dagger} \, e^{ i \vec{k} \cdot \vec{x}}\biggr].
\label{FOURb}
\end{eqnarray}
In Eqs. (\ref{FOURa})--(\ref{FOURb}) $\lambda= \oplus,\, \otimes$ runs over the tensor polarizations but, as in the conventional situation, the evolution of the mode functions is the same for each of the two values of $\lambda$:
\begin{equation}
\dot{G}_{k,\,\lambda} = - k^2 \,b^2 \, F_{k,\,\lambda}, \qquad \dot{F}_{k, \, \lambda} = \frac{G_{k,\, \lambda}}{b^2},
\label{FIVE}
\end{equation}
where we recall, as already mentioned in connection with Fig. \ref{FIG1} that 
the overdot now denotes a derivation with respect to $\eta$ (i.e. $\dot{F}_{k, \, \lambda} = 
\partial_{\eta} \,F_{k, \, \lambda}$).  From Eqs. (\ref{FOURa})--(\ref{FOURb}) the canonical commutation relations at equal $\eta$-times become 
\begin{equation}
[ \widehat{h}_{i\,j}(\vec{x}, \eta), \, \widehat{\pi}_{m\,n}(\vec{y}, \eta)] = i \, {\mathcal S}_{i\,j\,m\,n}(\hat{k}) \, \delta^{(3)}(\vec{x}- \vec{y}),
\label{SIX}
\end{equation}
In Eq. (\ref{SIX}) ${\mathcal S}_{i\,j\,m\,n}(\hat{k})$ follows from the sum over the tensor polarizations and the Wronskian of the mode functions is fixed:
\begin{eqnarray}
&& {\mathcal S}_{i\,j\,m\,n}(\hat{k}) = \frac{1}{4} \biggl[ p_{i\,m}(\hat{k})\, p_{j\,n}(\hat{k}) + p_{j\,m}(\hat{k})\, p_{i\,n}(\hat{k}) - p_{i\,j}(\hat{k})\, p_{m\,n}(\hat{k})\biggr], 
\label{SEVENa}\\
&& F_{k,\, \lambda}(\eta)\, G_{k,\, \lambda}^{\ast}(\eta) - F_{k,\, \lambda}^{\ast}(\eta)\, G_{k,\, \lambda}(\eta) \, = \, i/b^2(\eta),
\label{SEVENb}
\end{eqnarray}
where $p_{i \, j}(\hat{k}) = (\delta_{i\, j} - \hat{k}_{i}\, \hat{k}_{j})$ denotes the standard projector and $\hat{k}^{i} = k^{i}/|\vec{k}|$. Note that if the normalization of the Wronskian differs 
from Eq. (\ref{SEVENb}) the commutation relation of Eq. (\ref{SIX}) is not canonical 
anymore. 

The mode functions are normalized during the refractive phase where, according to Eq. (\ref{NEX}), the $\eta$-time and the conformal time are related as $(- \eta/\eta_{\ast}) = 
(- \tau/\tau_{\ast})^{\alpha/(1- \epsilon) +1}$ and $\eta_{\ast} = \tau_{\ast} (1 - \epsilon)/[n_{\ast} 
(1 + \alpha -\epsilon)]$; both relations follow from the definition of the 
$\eta$-time (i.e. $n(\eta) d\eta =d\tau$) and also from 
$n(a) = n_{*} (a/a_{\ast})^{\alpha}$, as implied by Eq. (\ref{NEX}) when $a< a_{\ast}$. In the $\eta$-time the explicit expression of $b(\eta)$ is: 
\begin{equation}
b(\eta) = b_{\ast} \biggl(-\frac{\eta}{\eta_{\ast}}\biggr)^{-\sigma}, \qquad 
\sigma= \frac{2- \alpha}{2( 1 + \alpha - \epsilon)}, \qquad \mathrm{for}\qquad \eta < - \eta_{\ast},
\label{EIGHT1}
\end{equation}
where $b_{\ast} = a_{\ast}/\sqrt{n_{\ast}}$. In the refractive regime the solution of Eq. 
(\ref{FIVE}) is therefore given by:
\begin{equation}
F_{k}(\eta) = \frac{{\mathcal N}}{\sqrt{ 2 k} \, b(\eta)} \, \sqrt{- k \eta} \, H_{\mu}^{(1)}(- k\, \eta), \qquad \qquad G_{k}(\eta) = - \frac{{\mathcal N}}{b(\eta)}\,\sqrt{\frac{k}{2}}  \,\sqrt{- k\eta} \, H_{\mu-1}^{(1)}(- k \eta), 
\label{EIGHT2}
\end{equation}
where $\mu = \sigma +1/2$ and $H_{\mu}^{(1)}(-k\eta)$ is the Hankel function of first kind \cite{abr1}; note that $|{\mathcal N}| = \sqrt{\pi/2}$ and, with 
this choice, the Wronskian normalization condition of Eq. (\ref{SEVENb}) 
is satisfied.  Thanks to Eq. (\ref{EIGHT2}) we can compute the tensor power spectrum that eventually determines the low-frequency normalization of the spectral energy density:
\begin{eqnarray}
{\mathcal P}_{T}(k, \eta) &=& \frac{4 \ell_{P}^2 \, k^{3}}{\pi^2} \bigl| F_{k}(\eta) \bigr|^2 \to
\biggl(\frac{H_{1}}{M_{P}}\biggr)^2\,\,{\mathcal C}(n_{T}, N_{*}, N_{t}, \epsilon) \,\, \biggl(\frac{k}{a_{1} H_{1}}\biggr)^{n_{T}},
 \label{EIGHT3}\\
 \qquad {\mathcal C}(n_{T}, N_{*}, N_{t}, \epsilon) &=& \frac{ 2^{6 - n_{T}}}{\pi^2} \biggl| 1 + \frac{\alpha}{1 -\epsilon}\biggr|^{2 - n_{T}}\, \Gamma^2\biggl(\frac{3 -n_{T}}{2}\biggr)e^{\alpha\, N_{\ast}( 3 - n_{T}) - n_{T} (N_{\ast} - N_{t})},
 \label{EIGHT3a}
 \end{eqnarray}
The second expression at the right hand side of Eq. (\ref{EIGHT3}) corresponds 
to the limit $| k \eta| \ll 1$ where ${\mathcal P}_{T}(k, \eta)$ becomes constant in time and the tensor to scalar ratio is:
\begin{equation}
r_{T}(\nu) = \frac{\epsilon}{\pi} \,  {\mathcal C}(n_{T}, N_{*}, N_{t}, \epsilon) \biggl(\frac{\nu}{\nu_{max}}\biggr)^{n_{T}}, \qquad\qquad \lim_{\alpha\to 0} {\mathcal C}(n_{T}, N_{*}, N_{t}, \epsilon) = 16\pi.
\label{EIGHT5}
\end{equation}
For $\alpha\to 0$ we have that $r_{T}\to 16 \epsilon$ and the standard consistency 
relation is recovered. In the same limit the tensor spectral index goes to $ - 2\epsilon$: 
\begin{equation}
n_{T} = \frac{3 \alpha - 2 \epsilon}{1+ \alpha - \epsilon} = \frac{3 \alpha}{1 + \alpha} + 
 \frac{\epsilon (\alpha - 2)}{(1 + \alpha)^2} + {\mathcal O}(\epsilon^2).
\label{EIGHT4}
\end{equation}
Equation (\ref{EIGHT4}) defines the tensor spectral index in the intermediate 
frequency range where $\alpha$ is  larger than $\epsilon$ and this 
is why the exact result can always be expanded in the limit $\epsilon \ll 1$. 
The slope of the tensor 
power spectrum evaluated for the wavelengths larger than the Hubble radius 
coincides, after reentry, with the slope of spectral energy density before the knee:
\begin{equation}
h_{0}^2 \,\,\Omega_{\mathrm{gw}}(\nu,\tau_{0}) = {\mathcal T}^2(\nu, \nu_{eq}) \,\,r_{\mathrm{T}}(\nu_{p}) \,\,\biggl(\frac{\nu}{\nu_{p}}\biggr)^{n_{\mathrm{T}}}, \qquad \nu < \nu_{\ast},
\label{EIGHT6}
\end{equation}
where $\nu_{p} =  3.092\,\,\mathrm{aHz}$ and ${\mathcal T}(\nu, \nu_{eq}) $ 
is:
\begin{equation}
{\mathcal T}(\nu, \nu_{eq}) = 6.453\times 10^{-8} \sqrt{1 + c_{2}\biggl(\frac{\nu_{eq}}{\nu}\biggr) + b_{2}\biggl(\frac{\nu_{eq}}{\nu}\biggr)^2},\qquad c_{2}= 0.5238, \qquad b_{2}=0.3537,
\label{EIGHT7}
\end{equation}
where the constants $c_{2}$ and $b_{2}$ follow from the 
transfer function for the spectral energy density \cite{TF} and are obtained 
 by integrating the mode functions across the radiation-matter transition and by computing 
$\Omega_{\mathrm{gw}}(\nu,\tau)$ for different frequencies. 
Above $\nu_{\ast}$ the spectral energy density is instead quasi-flat, as anticipated in Fig. \ref{FIG1} 
and it is approximately given by:
\begin{equation}
h_{0}^2 \Omega_{\mathrm{gw}}(\nu,\tau_{0}) = {\mathcal T}^2(\nu, \nu_{eq})\,\,r_{\mathrm{T}}(\nu_{p}) \,\,\biggl(\frac{\nu_{\ast}}{\nu_{p}}\biggr)^{n_{\mathrm{T}}}\,\,\biggl(\frac{\nu}{\nu_{\ast}}\biggr)^{- 2 \epsilon}\,  \qquad \nu_{\ast} < \nu < \nu_{max},
\label{EIGHT8}
\end{equation}
where the contribution of ${\mathcal T}^2(\nu, \nu_{eq})$ is, in practice, frequency-independent for 
$\nu > \nu_{\ast}$ and it can be approximated as $4.165\times 10^{-15}$. 
The results of Eqs. (\ref{EIGHT6})--(\ref{EIGHT7}) and (\ref{EIGHT8}) agree with the previous 
results of Ref. \cite{CC2} and are a natural candidate for exploring a
potential signal in the nHz range: the blue spectrum in the intermediate range and the flat slope 
at high-frequencies have a clear dynamical rationale which is encoded in Fig. \ref{FIG1}. 
It is however clear that the spectral energy density obtained in this concrete scenario does not support the 
 viewpoint that $n_{T}$, $r_{T}$ and $\nu_{\ast}$ could be independently  assigned without worrying about their mutual interplay. 

There are two classes of recent constraints that are relevant for the present discussions.  The $12.5$ yrs data set of the NANOGrav  pulsar timing array (PTA) experiment reported 
evidence of a stochastic signal in the nHz range. The features of this purported signal would imply, in the present notations, that\footnote{See also Ref. \cite{NANO2} for the $11$ yrs and Ref. \cite{NANO3} for the $9$ yrs data set where 
some useful upper limits on the relic gravitons have been reported.}\cite{NANO1}: 
\begin{equation}
10^{-8.6} < \, h_{0}^2 \Omega_{gw}(\nu,\tau_{0}) < \, 10^{-9.8}, \qquad\qquad 3\,\mathrm{nHz} < \nu< 100 \, \mathrm{nHz}.
\label{CONS1}
\end{equation}
The Kagra, LIGO and Virgo collaborations, in their attempt to constraint the stochastic backgrounds of relic gravitons 
also reported a constraint \cite{LIGO1} implying, in the case of a flat spectral energy density,
\begin{equation}
\Omega_{gw}(\nu, \tau_{0}) < 5.8 \times 10^{-9}, \qquad\qquad 20 \,\, \mathrm{Hz} < \nu_{KLV} < 76.6 \,\, \mathrm{Hz},
\label{CONS2}
\end{equation}
where $\nu_{KLV}$ denotes the Kagra-LIGO-Virgo frequency. 
This limit improves on a series of bounds previously deduced by the wide-band interferometers (see \cite{CC0} for a recent review). In particular 
in Ref. \cite{LIGO2} the analog of Eq. (\ref{CONS2}) implied $ \Omega_{gw}(\nu, \tau_{0}) < 6 \times 10^{-8}$ 
for a comparable frequency interval and always in the case of a flat spectral energy density. While the bound of Eq. (\ref{CONS2}) could be immediately 
used also in our case since at high-frequency the spectral energy density is 
nearly scale-invariant, it is useful to elaborate on the result of Ref. \cite{LIGO1}. 
The collaboration actually reports a threefold bound which could be parametrized as 
\begin{equation}
\Omega_{gw}(\nu, \tau_{0}) = \overline{\Omega}(\delta) \biggl(\frac{\nu}{\nu_{ref}}\biggr)^{\delta}, \qquad \qquad \nu_{ref} = 25 \,\, \mathrm{Hz}.
\label{NOT1}
\end{equation}
In terms of Eq. (\ref{NOT1}) the results of Ref. \cite{LIGO1} read 
$\overline{\Omega}(0) < 5.8 \times 10^{-9}$ (valid in the case $\delta =0$),
$\overline{\Omega}(2/3) < 3.4 \times 10^{-9}$ (when $\delta= 2/3$) 
and $\overline{\Omega}(3) < 3.9 \times 10^{-10}$ (when $\delta= 3$).
When the value of $\delta$ increases the bound becomes more restrictive 
once the reference frequency has been kept fixed. The three results are unified in the following interpolating formula 
\begin{equation}
\log{\overline{\Omega}}(\delta) < -\,8.236 -\, 0.335\, \delta- 0.018\, \delta^2.
\label{NOT2}
\end{equation}
While the quadratic fit is slightly more accurate the linear one, 
the essence of the arguments does not change in the two cases since the different points fall approximately on the same straight line (i.e. $ - 8.223 - 0.393\, \delta$). Since
in the present case the bound (\ref{NOT2}) should be applied at high-frequencies we will have $ \delta =- 2 \epsilon/ (1- \epsilon)$ with $\epsilon < 0.01$.
We can immediately see from Eq. (\ref{NOT2}) that, to leading order in $\epsilon$, 
Eq. (\ref{NOT2}) implies that $\log{\overline{\Omega}}(\epsilon) < -\,8.236 -\, 0.335\, \epsilon -0.393 \epsilon^2$ giving, at most,  a negligible correction in the exponent.
The pivotal parameters that determine the spectrum are  $\alpha$, $N_{\ast}$ and $N_{t}$.
If $N_{*}$ is of the order of $N_{t}$ the transition to normalcy occurs at the end of inflation\footnote{
It is not difficult to show that in this case, as previously discussed, it is impossible to get a 
large signal in the nHz range without jeopardizing the big-bang nuclosynthesis constraint
(see \cite{CC2,CC3}). This point is quite well known also from other 
scenarios where the blue spectra are not induced by the variation of the refractive index \cite{CC0}.
It is clear that this conclusion holds, in the present context, because $\nu_{max}$ is cannot be arbitrarily fixed.}
When $N_{*} <N_{t}$ the transition to normalcy takes place well before the onset of the radiation-dominated epoch (i.e. when the background is still inflating deep inside the quasi-de Sitter stage of expansion). The scale-dependent constraint on the tensor to scalar ratio is conventionally applied at the pivot scale $k_{p}$ (see the discussion after Eq. (\ref{NEX2})) corresponding to the comoving frequency $\nu_{p} = {\mathcal O}(3)$ aHz. In a conservative perspective we therefore require the limit \cite{RT1,RT2} $r_{T}(\nu_{p}) \leq 0.06$.
The bounds coming from big-bang nucleosynthesis \cite{bbn1,bbn2,bbn3} imply instead a constraint on the integral of the spectral energy density of the gravitons:
\begin{equation}
h_{0}^2  \int_{\nu_{bbn}}^{\nu_{max}}
  \Omega_{\mathrm{gw}}(\nu,\tau_{0}) d\ln{\nu} = 5.61 \times 10^{-6} \Delta N_{\nu} 
  \biggl(\frac{h_{0}^2 \Omega_{\gamma0}}{2.47 \times 10^{-5}}\biggr),
\label{CC2}
\end{equation}
where $\Omega_{\gamma0}$ is the (present) critical fraction of CMB photons.
The lower limit of integration in Eq. (\ref{CC2}) is given by  the frequency corresponding to the Hubble rate at the nucleosynthesis epoch: 
\begin{equation}
\nu_{bbn}= 2.252\times 10^{-11} \biggl(\frac{N_{eff}}{10.75}\biggr)^{1/4} \biggl(\frac{T_{\mathrm{bbn}}}{\,\,\mathrm{MeV}}\biggr) 
\biggl(\frac{h_{0}^2 \Omega_{\mathrm{R}0}}{4.15 \times 10^{-5}}\biggr)^{1/4}\,\,\mathrm{Hz} \simeq 0.01\, \mathrm{nHz},
\label{CC3}
\end{equation}
where  $N_{eff}$ denotes the effective number of relativistic degrees of freedom entering the total energy density of the plasma and $T_{\mathrm{bbn}}$ is the temperature of big-bang nucleosynthesis. The limit  of Eq. (\ref{CC2}) sets an indirect constraint  on the extra-relativistic species (and, among others, on the relic gravitons). Since Eq. (\ref{CC2}) is relevant in the context of neutrino physics (when applied to massless fermionic species), the limit is often expressed for practical reasons  in terms of $\Delta N_{\nu}$ representing the contribution of supplementary neutrino species. The actual bounds on $\Delta N_{\nu}$ range from $\Delta N_{\nu} \leq 0.2$ 
to $\Delta N_{\nu} \leq 1$;  the integrated spectral density in Eq. (\ref{CC2}) is thus between $10^{-6}$ and $10^{-5}$. 

\begin{figure}[!ht]
\centering
\includegraphics[height=7cm]{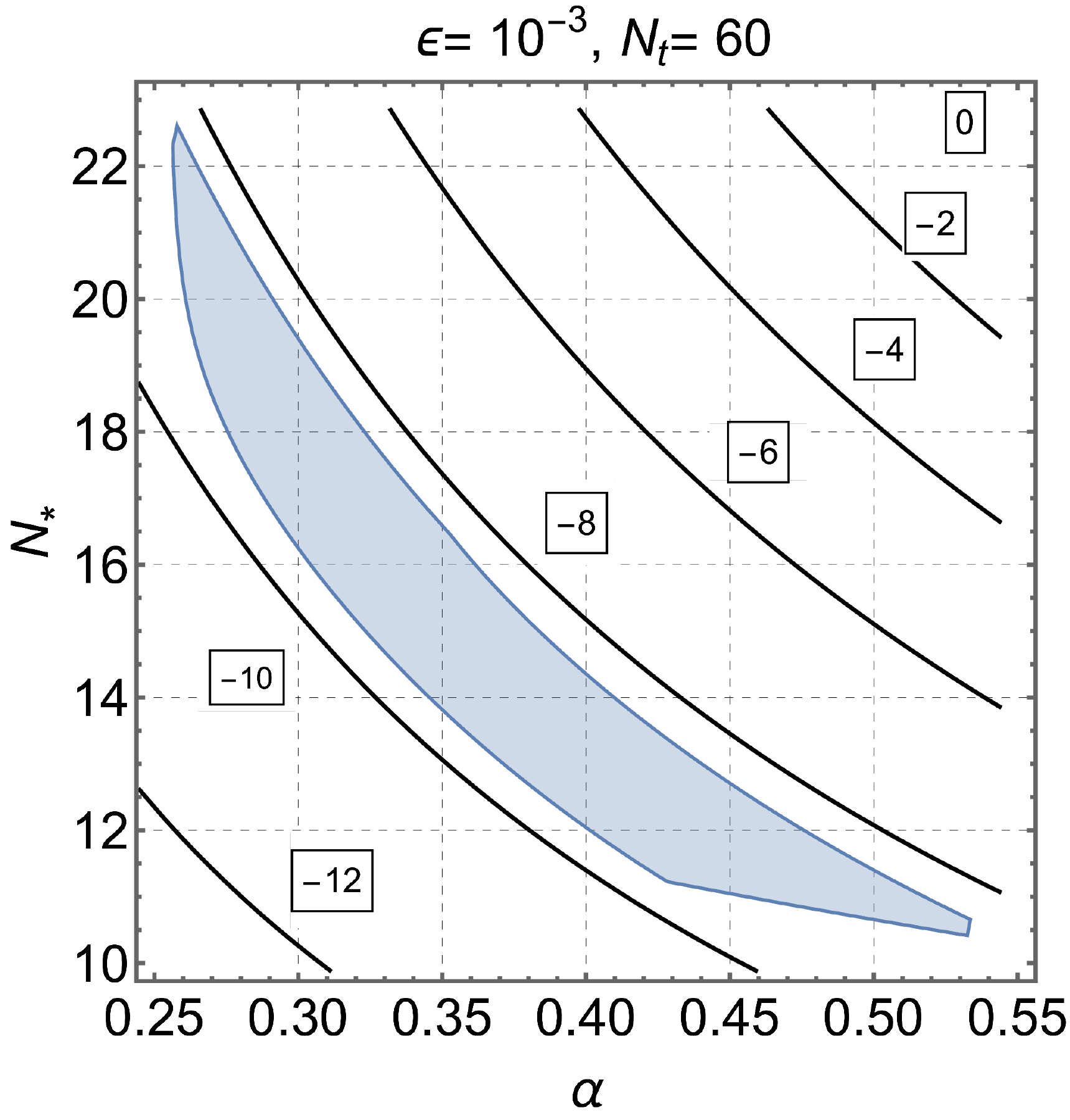}
\includegraphics[height=7cm]{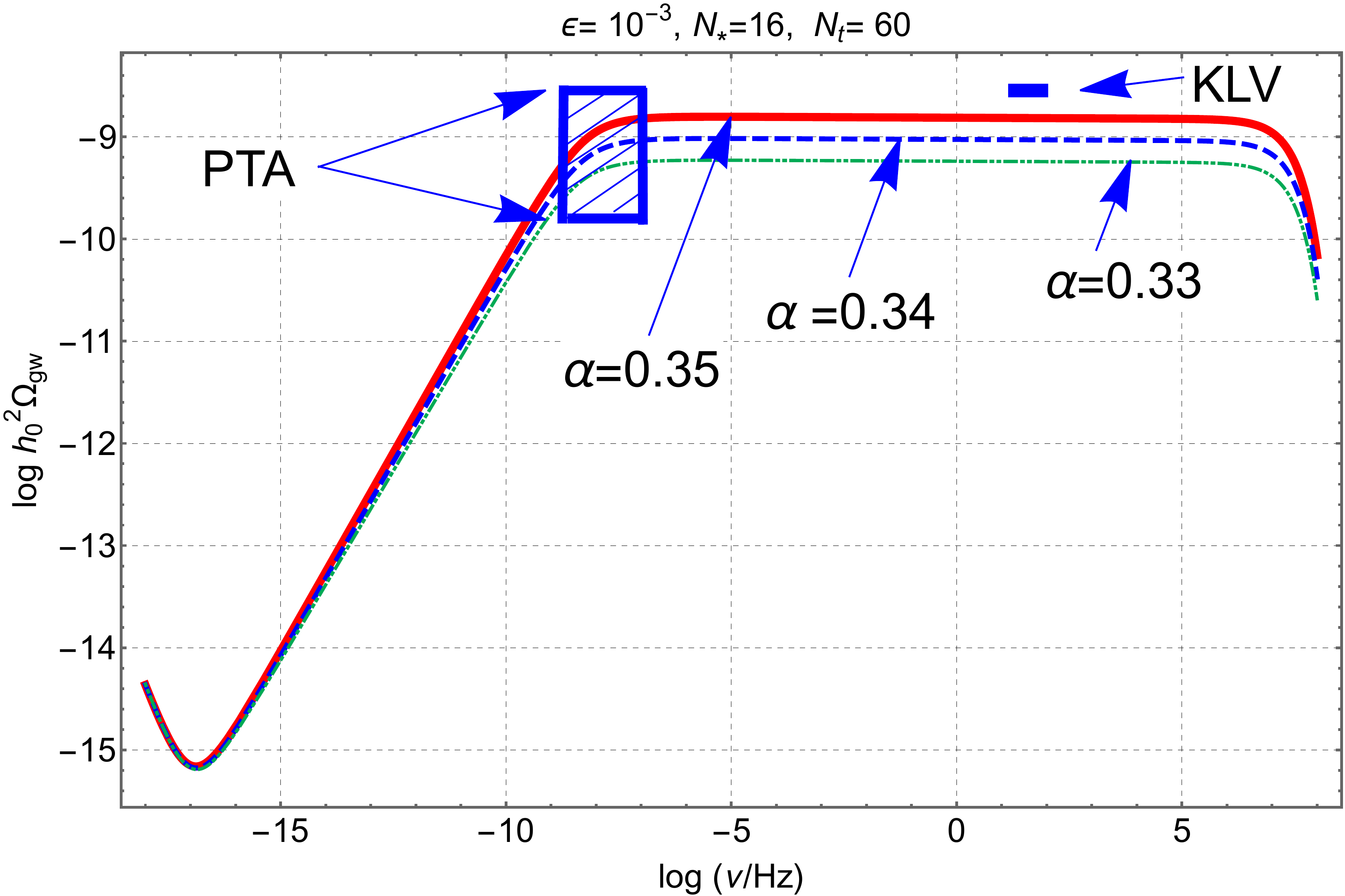}
\caption[a]{The allowed region of the parameter space is illustrated together with the explicit form of the spectral energy density 
when the corresponding parameters are selected within the shaded region where all the relevant are satisfied.
 The different contours 
appearing in the plot at the left correspond to the values of $h_{0}^2 \, \Omega_{gw}(\nu_{KLV}, \tau_{0})$ 
where $\nu_{KLV}$ approximately denotes most sensitive frequency domain  
of the wide-band interferometers (i.e. Kara-LIGO-Virgo) that recently provided interesting results that can be interpreted as upper limits on the spectral energy density of the relic gravitons. The shaded box highlights the region suggested by the PTA. We finally recall that $\log{}$ denotes the common logarithm (as opposed to 
the natural logarithm) employed, for instance, in the definition of the number of $e$-folds (see Eq. (\ref{NEX3}) 
and comments thereafter).}
\label{FIG2}      
\end{figure}

The limits stemming from CMB analyses (i.e. $r_{T}(\nu_{p}) < 0.06$), the requirement of Eq. (\ref{CONS1}), 
the Kagra-LIGO-Virgo limit (KLV) of Eq. (\ref{CONS2}) (see also Eqs. (\ref{NOT1})--(\ref{NOT2})) and the BBN constraints of Eq. (\ref{CC2}) have all been enforced in Fig. \ref{FIG2}. The shaded area (appearing in the plot at the left of Fig. \ref{FIG2} and resembling a circular segment) illustrates the region where are all the various constraints are concurrently satisfied when $N_{t}$ is fixed to a fiducial value. On top of the mentioned bounds, for consistency it is preferable to require,  in a model-independent perspective, that $\nu_{\ast} > \nu_{bbn}$: if this condition is not included the allowed region in Fig. \ref{FIG2} gets even larger.
The shaded box appearing in Fig. \ref{FIG2} illustrates the region defined by Eq. (\ref{CONS1}) where the tentative signal of the PTA should be approximately located. In the right plot of Fig. \ref{FIG2} the spectral energy density is reported for different values of $\alpha$ and $N_{\ast}$ all falling within the shaded region that appears in the left plot. When the parameters are selected within the circular segment the corresponding spectral energy density falls within the PTA box while the limits coming from wide-band interferometers are satisfied.  The results 
of Fig. \ref{FIG2} suggest that a large signal in the nHz range is obtained when the variation 
of the refractive index occurs sufficiently early during the inflationary stage and anyway not 
beyond the first $20$ $e$-folds. In units of the inflationary Hubble rate, the rate of variation of the refractive index must fall in the range $0.25 < \alpha < 0.55$. For this portion of the parameter space the frequency of the knee 
is in the nHz range. 

\begin{figure}[!ht]
\centering
\includegraphics[height=7cm]{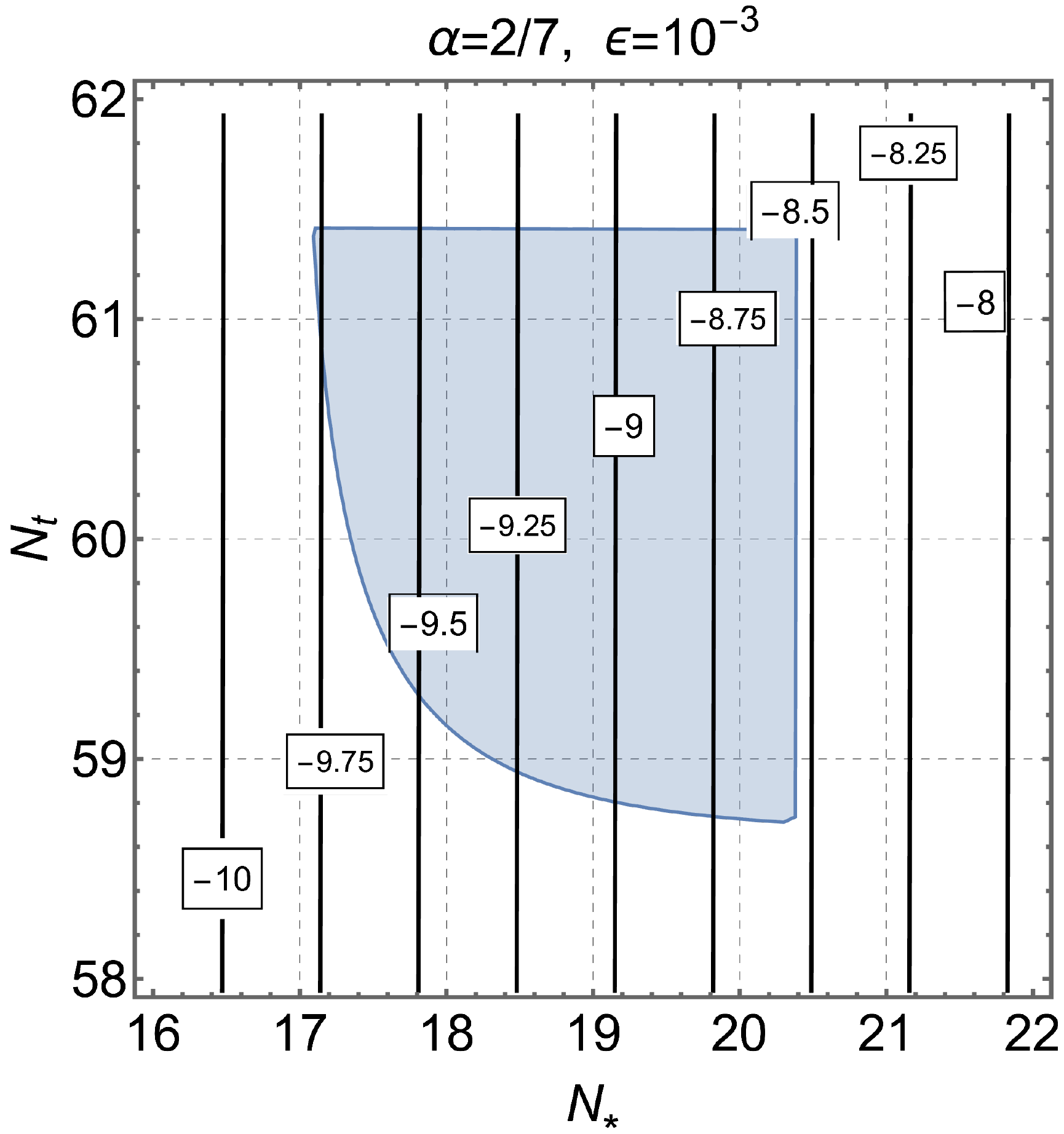}
\includegraphics[height=7cm]{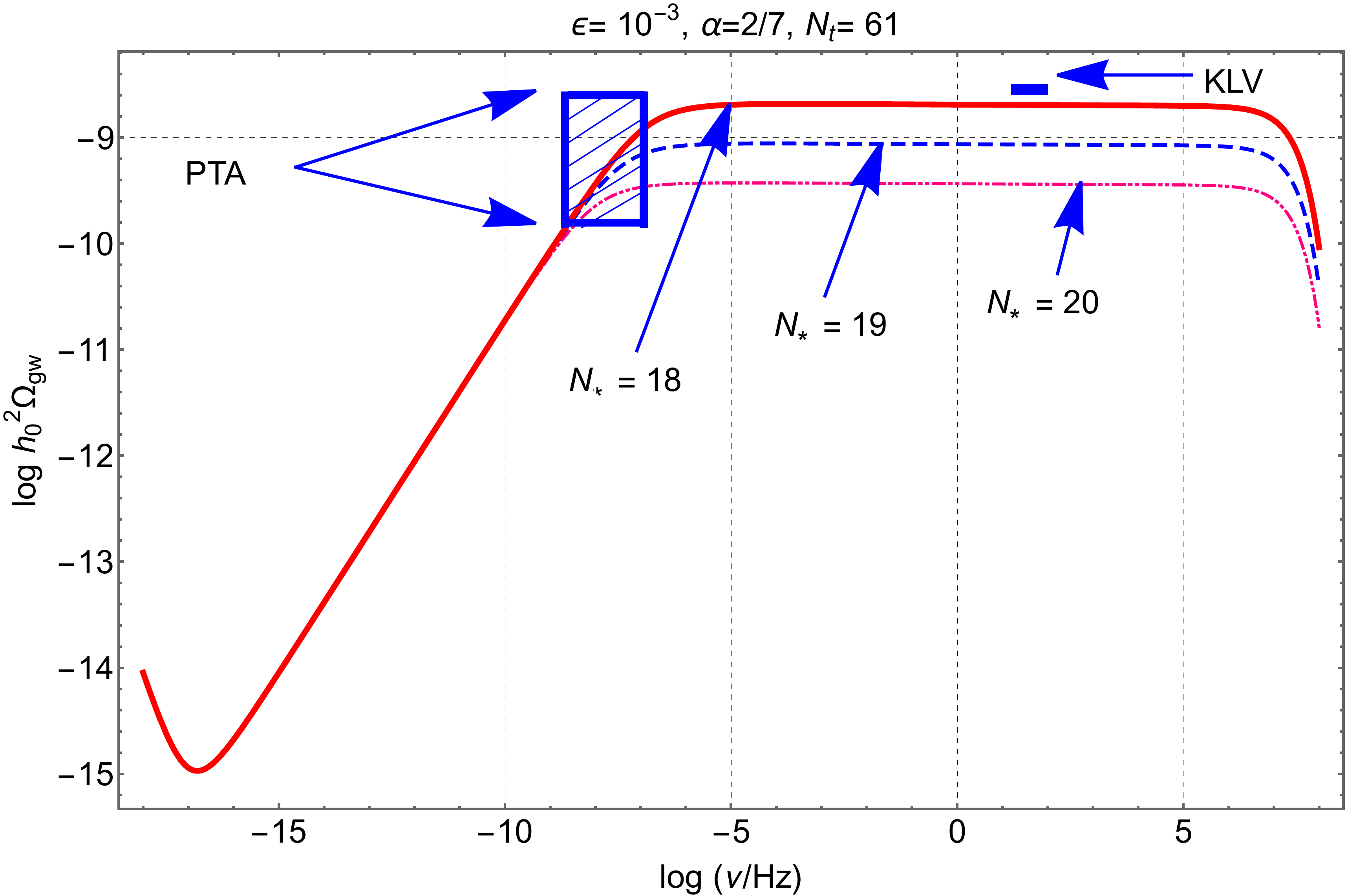}
\caption[a]{In the plot at the left we compute the allowed region in the parameter 
space when $\alpha$ is fixed to $2/7$. This choice implies that the chirp amplitude 
scales as $\nu^{-2/3}$ while the spectral energy density goes as $\nu^{2/3}$. In the plot at the 
right we illustrate the spectral energy density for a set of parameters selected 
within the shaded region appearing in the exclusion plot at the right. 
As in Fig. \ref{FIG2} common logarithms are employed on both axes of the right plot.}
\label{FIG3}      
\end{figure}

It is now interesting to fix  $\alpha$ to some fiducial value and to deduce the constraints in the plane $(N_{\ast}, \, N_{t})$: this strategy is illustrated in Fig. \ref{FIG3}. It is 
clear that for $\alpha\simeq 2/7$ the allowed region and the maximal signal of the model 
(see the right plot of Fig. \ref{FIG3}) occur for $N_{t} = {\mathcal O}(60)$ and $N_{\ast} = {\mathcal O}(20)$.  
Furthermore for $ \alpha = 2/7 \simeq  {\mathcal O}(0.28)$, $N_{t}= {\mathcal O}(60)$ 
and $N_{\ast} = {\mathcal O}(20)$ the spectral energy density passes through the PTA box and undershoots 
the $KLV$ bound.  The rationale for the apparently peculiar choice of $\alpha$ appearing in Fig. \ref{FIG3} stems, in short, from the following consideration. The PTA results are often reported in terms of a chirp amplitude scaling as $\nu^{-2/3}$ for a typical reference frequency ${\mathcal O}(\mathrm{yr}^{-1})$. If, at the present time, $h_{c}(\nu,\tau_{0})$ scales as $\nu^{\beta}$ the spectral energy density has the following dependence on the comoving frequency\footnote{See Ref. \cite{CC0} for a discussion of the 
chirp amplitude and of its relation with the spectral energy density. In general terms the chirp amplitude 
is simply related to the tensor power spectrum as $2 h_{c}^2(k,\eta) = {\mathcal P}_{T}(k,\eta)$. }:
\begin{equation}
h_{c}(\nu,\tau_{0}) = Q \biggl(\frac{\nu}{\nu_{ref}}\biggr)^{\beta}, \qquad \Rightarrow \qquad  \Omega_{gw}(\nu,\tau_{0}) = \frac{2\, \pi^2 \, \nu^2}{3 H_{0}^2}  Q^2 \biggl(\frac{\nu}{\nu_{ref}}\biggr)^{2\alpha} \propto \biggl(\frac{\nu}{\nu_{ref}}\biggr)^{2+ 2\beta},
\label{NANOeq1}
\end{equation}
where, as already mentioned, $\nu_{ref} = {\mathcal O}(\mathrm{yr}^{-1})$ and  $Q$ just denotes a constant amplitude. If $\beta = -2/3$ we have that $\Omega_{gw} \propto \nu^{2/3}$. The value of $\alpha$ corresponding to $\beta= -2/3$ is obtained by setting $n_{T} \simeq 2/3$ in Eq. (\ref{EIGHT4}). 
Consequently we have that $\alpha = (2 + 4 \epsilon)/7$ 
which can be approximated as $\alpha = 2/7 + {\mathcal O}(\epsilon)$ 
since $\epsilon < 10^{-3}$. In this case the spectral energy density  
is illustrated in the right plot of Fig. \ref{FIG3} for a set of values
selected within the allowed region of the parameter space. If we would really pretend that the PTA measures a real signal with chirp amplitude proportional to $\nu^{-2/3}$ the region of the parameter space illustrated in Fig. \ref{FIG3} should be the most promising one. 

\begin{figure}[!ht]
\centering
\includegraphics[height=6.5cm]{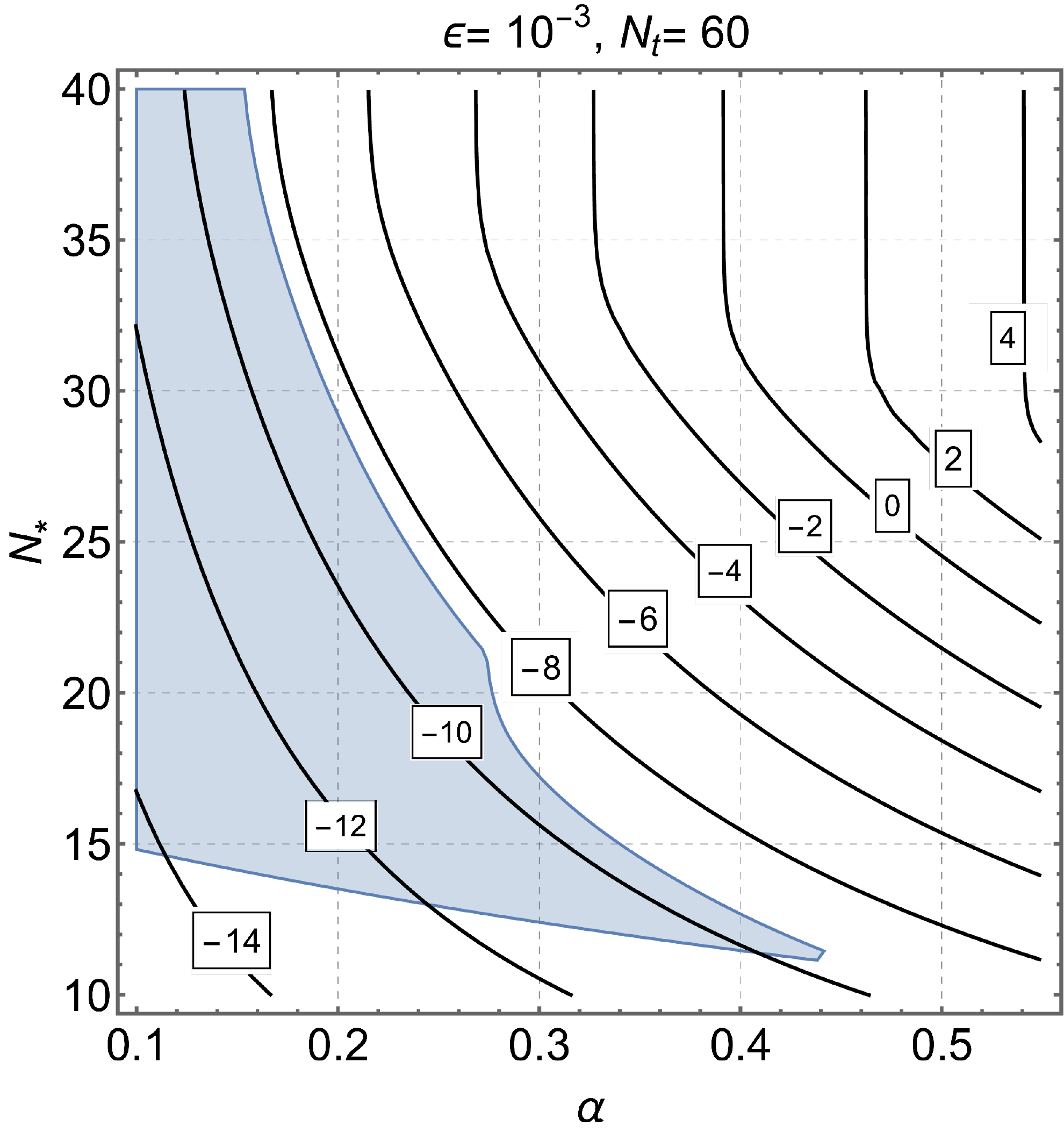}
\includegraphics[height=6.5cm]{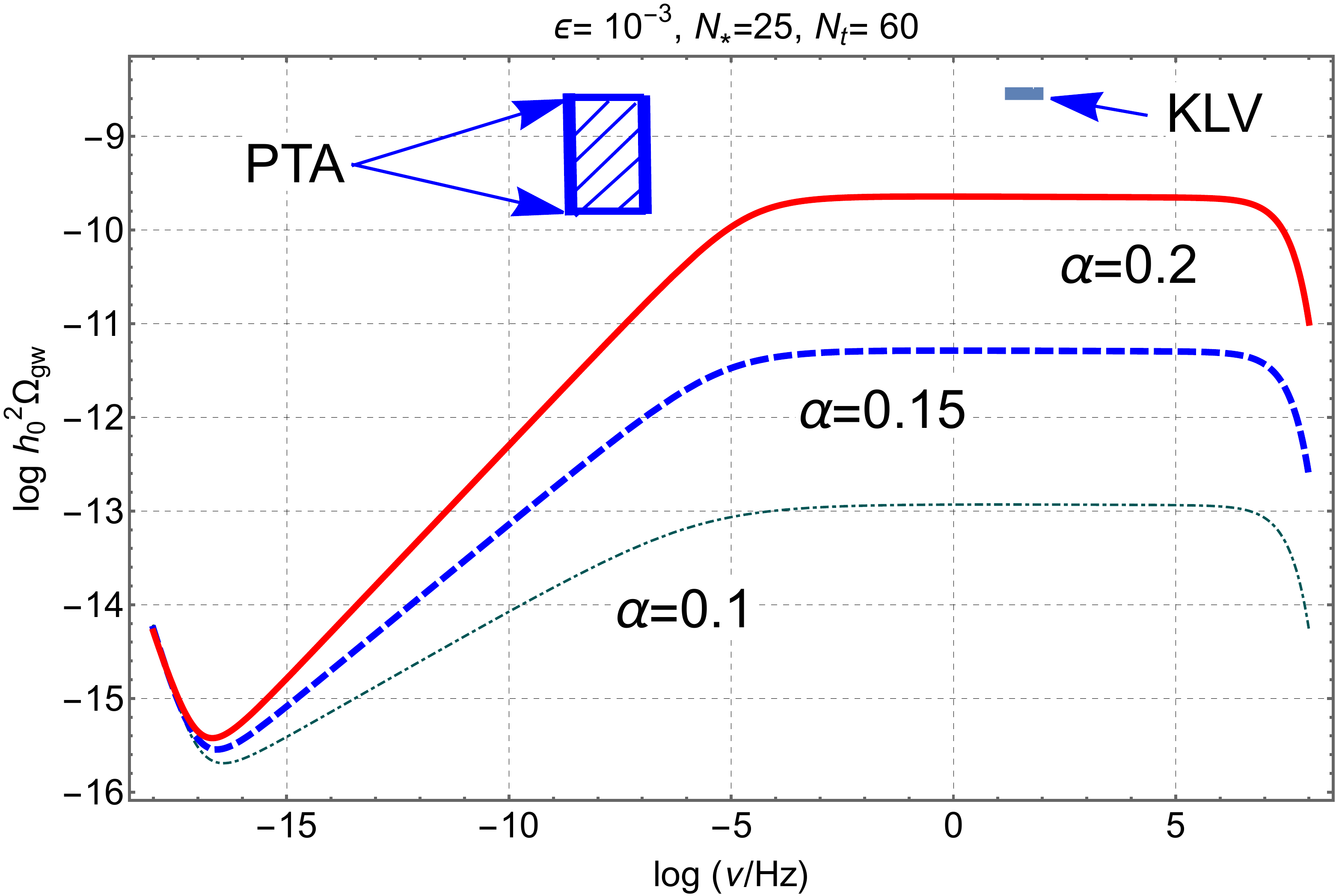}
\caption[a]{The shaded area is obtained by imposing all the previous constraints except 
for Eq. (\ref{CONS1}) which is instead replaced by the limit $h_{0}^2 \Omega_{gw}(\nu_{ref}, \tau_{0}) < 3.4(1) \times 10^{-10}$ for $\nu_{ref} = {\mathcal O}(10^{-8})$ Hz. In the plot 
at the right a specific choice of the parameters. This result shows that if the PTA results are interpreted as 
upper limits (as originally done when introducing the variation of the refractive index \cite{CC0}) 
the range of variation of $N_{*}$ and $\alpha$ gets larger even in the minimal model examined here.}
\label{FIG4}      
\end{figure}

If the results of the PTA are interpreted as an upper limit,  the allowed region of the parameter space gets wider as we can argue by comparing the left plot of Fig. \ref{FIG4} with the corresponding result reported in Fig. \ref{FIG2}, always in the plot at the left. The logic of the upper limit has been encouraged by the observational collaborations themselves in the 11-yr release \cite{NANO2} where the PTA data implied, within our notations, that $h_{0}^2 \Omega_{gw}(\nu_{ref}, \tau_{0}) < 3.4(1) \times 10^{-10}$.
If we apply this viewpoint to the present case (and consequently avoid the requirements associated with 
Eq. (\ref{CONS1})) we obtain the exclusion plot reported at the left of Fig. \ref{FIG4} where
the shaded area corresponds, as usual, to the allowed region. From Fig. \ref{FIG4}
 the area where $\alpha < 0.25$ is now viable and it corresponds to a value of $N_{\ast}$ 
that is comparatively larger than in Fig. \ref{FIG2}. It is however clear that, in this region, 
the purported ``signal'' attributed to relic gravitons cannot be reproduced. Indeed in the right plot of Fig. 
\ref{FIG4} the spectral energy density of the relic gravitons always undershoots the PTA box and 
the upper limits of wide-band detectors. 

All in all the evolution of refractive index of the gravitons 
during the early stages of the inflationary evolution leads to a blue (i.e. slightly increasing) slope
of the spectral energy density at intermediate frequencies above the fHz. 
While the specific values of the slopes are determined by the competition of the slow-roll parameter and of the 
rate of variation of the refractive index, the general idea suggests that increasing frequency spectra 
can also be obtained in the framework of conventional inflationary 
scenarios.  Some time ago  \cite{CC0}  (see also \cite{CC3,CC5}) it was suggested
that the allowed regions of the parameter space could only be probed 
by some of the planned detectors operating in the audio band or in the 
mHz region. Broadly speaking the previous results assumed  
$h_{0}^2 \Omega_{gw}(\nu_{pulsar}, \tau_{0}) < 10^{-8}$ 
for $\nu_{pulsar}= {\mathcal O}(10)$ nHz. The generous bounds previously applied 
(and motivated by the early stage of the PTA) 
are now being improved. In this investigation, following 
some recent claims, we focussed the attention on the nHz domain
and demonstrated that the recent findings, if confirmed, are 
compatible with the variation of the refractive index of the
relic gravitons during the early stages of the inflationary evolution. The results reported here have been purposely derived in the minimal framework where the high-frequency plateau ranges between the nHz 
and the MHz region. If the slow-roll condition is violated towards the end of inflation the spectral energy density is not constant at high-frequencies and the flat plateau is replaced by a decreasing branch. In this case the spectral energy density may even develop a maximum between the nHz and the 
kHz. To address a large signal in the nHz domain the present analysis suggests that the realistic spectra should have at least three branches: the low-frequency branch in the aHz region originally discussed in \cite{AA3}, an intermediate region where the spectral index increases and the high-frequency domain (extending up to the MHz) where the spectrum is nearly scale-invariant. We could easily have spectra with four different branches \cite{CC4}. This behaviour arises since the post-inflationary evolution is not dominated by radiation but it contains either a stiff epoch or a delayed reheating. 

The inflationary evolution of the refractive has been discussed in a minimal scenario where the spectral energy density of cosmic gravitons is blue at intermediate frequencies and then flattens out at high-frequency. The obtained spectra satisfy all the constraints associated with the stochastic 
backgrounds of relic gravitons and may also explain the recent findings of the PTA that however need a convincing confirmation. While the EPTA (European Pulsar Timing Array) \cite{EPTA1,EPTA2} and the PPTA  (Parkes Pulsar Timing Array) \cite{PPTA1,PPTA2} should eventually offer an independent (and convincing) evidence  for the 
claim, it would also be quite interesting to see if the present bounds coming from wide-band interferometers could be improved by, at least, a couple of orders of magnitude. The improved limits would allow to decide if the spectral energy-density at high-frequencies is really flat or should instead decrease.  The theoretical perspective conveyed here suggests that the blue spectral index, the tensor to scalar ratio and the knee of the spectra cannot be independently assigned but their mutual relations are essential to determine the allowed regions of the parameter space. For this 
reason we expect that the potential signatures of the refractive index will be distinguishable 
from other potential sources of power at intermediate frequencies. If the forthcoming 
measurements will not support the evidences currently claimed by the PTA measurements 
we will have to conclude that the variation of the refractive index may be even longer and potentially lead to a 
large signal in the MHz domain. In this case, however, also other spikes are potentially expected 
 and the improved bounds of the PTA arrays (together with the upper limits of the 
wide-band detectors) will clarify the origin of the signals in the high-frequency 
and ultra-high-frequency domains.

The author wishes to thank T. Basaglia, A. Gentil-Beccot, S. Rohr and J. Vigen of the CERN Scientific Information Service for their valuable help.


\newpage

\begin{thebibliography}{99}

\itemsep -3pt

\bibitem{AA1} L.~P.~Grishchuk,   Sov.\ Phys.\ JETP {\bf 40}, 409 (1975)   [Zh.\ Eksp.\ Teor.\ Fiz.\  {\bf 67}, 825 (1974)].

\bibitem{AA1a} L.~P.~Grishchuk,  Annals N.\ Y.\ Acad.\ Sci.\  {\bf 302}, 439 (1977).
  
\bibitem{AA2} A.~A.~Starobinsky, JETP Lett.\  {\bf 30}, 682 (1979) [Pisma Zh.\ Eksp.\ Teor.\ Fiz.\  {\bf 30}, 719 (1979)]. 

\bibitem{AA3} V. A. Rubakov, M. V. Sazhin and A. V. Veryaskin, Phys. Lett. B {\bf 115}, 189 (1982).
  
\bibitem{CC0}  M.~Giovannini, Prog. Part. Nucl. Phys. {\bf 112}, 103774 (2020).
  
\bibitem{CC1} P.~Szekeres,  Annals Phys.\  {\bf 64}, 599 (1971);  P.~C.~Peters,  Phys.\ Rev.\ D {\bf 9}, 2207 (1974).

\bibitem{CC2}  M.~Giovannini, Class.\ Quant.\ Grav.\  {\bf 33}, 125002 (2016)  [arXiv:1507.03456 [astro-ph.CO]].

\bibitem{ONEW} S. Weinberg, Phys. Rev. D {\bf 77}, 123541 (2008).

\bibitem{TWO}  S.-Y. Pi and R. Jackiw, Phys. Rev. D {\bf 68}, 104012 (2003).

\bibitem{THREE} A. Lue, L. Wang, and M. Kamionkowski, Phys. Rev. Lett. {\bf 83}, 1506 (1999).

\bibitem{FOUR} M.~Giovannini,  Phys. Rev. D {\bf 99},  083501 (2019).

\bibitem{NON1} H.~Motohashi and A.~A.~Starobinsky, JCAP {\bf 11}, 025 (2019).

\bibitem{NON2} M.~Guerrero, D.~Rubiera-Garcia and D.~Saez-Chillon Gomez, Phys. Rev. D {\bf 102}, 123528 (2020).

\bibitem{NON3} A.~Mohammadi, T.~Golanbari, S.~Nasri and K.~Saaidi, Phys. Rev. D {\bf 101},  123537 (2020).

\bibitem{NON4} M.~Gasperini and M.~Giovannini, Phys. Lett. B {\bf 287}, 56 (1992).

\bibitem{NON5} I.~Antoniadis, J.~Rizos and K.~Tamvakis, Nucl. Phys. B {\bf 415}, 497 (1994).

\bibitem{NON6} Z.~K.~Guo and D.~J.~Schwarz, Phys. Rev. D {\bf 80}, 063523 (2009).

\bibitem{CC3} M.~Giovannini, Eur. Phys. J. C {\bf 78}, 442 (2018).

\bibitem{CC5} M.~Giovannini, Phys. Rev. D {\bf 98}, 103509 (2018).

\bibitem{FIVEa} L.~H.~Ford and L.~Parker, Phys. Rev. D {\bf 16}, 1601 (1977).

\bibitem{TF} M.~Giovannini, Phys. Lett. B {\bf 668}, 44 (2008); Class. Quant. Grav. {\bf 26}, 045004 (2009).

\bibitem{RT1}  Y.~Akrami {\it et al.} [Planck Collaboration], Astron. Astrophys. {\bf 641}, A10 (2020).

\bibitem{RT2}  N.~Aghanim {\it et al.} [Planck Collaboration], Astron. Astrophys. {\bf 641}, A6 (2020).

\bibitem{abr1} M. Abramowitz and I.A. Stegun, {\it Handbook of Mathematical Functions} (Dover, New York, 1972).
 
\bibitem{NANO1} Z. Arzoumanian {\it et al.}, Astrophys. J. Lett. {\bf 905}, L34 (2020).

\bibitem{NANO2} Z. Arzoumanian {\it et al.}, Astrophys. J. {\bf 859}, 47 (2018).

\bibitem{NANO3} Z.~Arzoumanian {\it et al.} [NANOGrav], Astrophys. J. {\bf 821}, 13 (2016)

\bibitem{LIGO1} R.~Abbott {\it et al.} [KAGRA, Virgo and LIGO Scientific], Phys. Rev. D {\bf 104}, 022004 (2021).

\bibitem{LIGO2} B.~P.~Abbott {\it et al.} [LIGO Scientific and Virgo], Phys. Rev. D {\bf 100}, 061101 (2019).

\bibitem{bbn1} V.~F.~Schwartzmann, JETP Lett.\ {\bf 9}, 184 (1969).
  
\bibitem{bbn2} M.~Giovannini, H.~Kurki-Suonio and E.~Sihvola, Phys.\ Rev.\  D {\bf 66}, 043504 (2002).

\bibitem{bbn3} R.~H.~Cyburt, B.~D.~Fields, K.~A.~Olive, and E.~Skillman, Astropart.\ Phys.\ {\bf 23}, 313 (2005).

\bibitem{CC4} M.~Giovannini, Phys. Lett. B {\bf 789}, 502 (2019).

\bibitem{EPTA1} L. Lentati {\it et al.}, Mon. Not. Roy. Astron. Soc. {\bf 453},  2576 (2015).

\bibitem{EPTA2} G. Desvignes {\it et al.}, Mon. Not. Roy. Astron. Soc. {\bf 458}, 3341 (2016).

\bibitem{PPTA1} D.~R.~B.~Yardley, W.~A.~Coles, G.~B.~Hobbs, J.~P.~W.~Verbiest, R.~N.~Manchester {\it et al.}
Mon. Not. Roy. Astron. Soc. {\bf 414}, 1777 (2011).

\bibitem{PPTA2} B. B. P. Perera {\it et al.}, Mon. Not. Roy. Astron. Soc. {\bf 490},  4666 (2019).

\end{thebibliography}
\end{document}